\documentclass[twocolumn]{aastex631}

\newcommand{\dotdeg}{$\rlap{.}^\circ$}
\usepackage{graphicx}
\usepackage{hyperref}
\usepackage{multirow}

\graphicspath{{./}{figures/}}

\shorttitle{Observations of Uranus at High Phase Angle by New Horizons}
\shortauthors{Hasler et al.}

\begin{document}

\title{Observations of Uranus at High Phase Angle as Seen by New Horizons}

\correspondingauthor{S. N. Hasler}
\email{shasler@mit.edu}

\author[0000-0002-4894-193X]{Samantha N. Hasler}
\affiliation{Department of Earth, Atmospheric and Planetary Sciences, Massachusetts Institute of Technology, 77 Massachusetts Ave, Cambridge, MA 02139, USA}

\author[0000-0002-4321-4581]{L. C. Mayorga}
\affiliation{Johns Hopkins University, Applied Physics Laboratory, 11100 Johns Hopkins Road, Laurel, MD 20723, USA}

\author[0000-0002-8296-6540]{William M. Grundy}
\affiliation{Lowell Observatory, 1400 W. Mars Hill Road, Flagstaff, AZ 86001, USA}

\author[0000-0003-4641-6186]{Amy A. Simon}
\affiliation{NASA Goddard Space Flight Center, Greenbelt, MD 20771, USA}

\author[0000-0001-8821-5927]{Susan D. Benecchi}
\affiliation{Planetary Science Institute, 1700 East Fort Lowell, Suite 106
Tucson, AZ 85719}

\author[0000-0003-1869-4947]{Carly J. A. Howett}
\affiliation{Planetary Science Institute, 1700 East Fort Lowell, Suite 106
Tucson, AZ 85719}

\author[0000-0001-8541-8550]{Silvia Protopapa}
\affiliation{Southwest Research Institute, 1050 Walnut St., Suite 300 Boulder, CO 80302, USA}

\author[0000-0001-8751-3463]{Heidi B. Hammel}
\affiliation{Association of Universities for Research in Astronomy, 1331 Pennsylvania Ave. NW, Washington, DC 20004, USA}

\author[0009-0000-4203-7141]{Daniel D. Wenkert}
\affiliation{Jet Propulsion Laboratory, California Institute of Technology, 4800 Oak Grove Drive, Pasadena, CA 91109, USA}

\author[0000-0001-5018-7537]{S. Alan Stern}
\affiliation{Southwest Research Institute, 1050 Walnut St., Suite 300 Boulder, CO 80302, USA}

\author[0000-0003-3045-8445]{Kelsi N. Singer}
\affiliation{Southwest Research Institute, 1050 Walnut St., Suite 300 Boulder, CO 80302, USA}

\author[0000-0003-0333-6055]{Simon B. Porter}
\affiliation{Southwest Research Institute, 1050 Walnut St., Suite 300 Boulder, CO 80302, USA}

\author[0000-0002-4644-0306]{Pontus C. Brandt}
\affiliation{Johns Hopkins University, Applied Physics Laboratory, 11100 Johns Hopkins Road, Laurel, MD 20723, USA}

\author[0000-0002-3672-0603]{Joel W. Parker}
\affiliation{Southwest Research Institute, 1050 Walnut St., Suite 300 Boulder, CO 80302, USA}

\author[0000-0002-3323-9304]{Anne J. Verbiscer}
\affiliation{Department of Astronomy, University of Virginia, Charlottesville, VA 22904, USA}

\author{John R. Spencer}
\affiliation{Southwest Research Institute, 1050 Walnut St., Suite 300 Boulder, CO 80302, USA}

\author{the New Horizons Planetary Science Theme Team}

\begin{abstract}

We present flux measurements of Uranus observed at phase angles of 43.9°, 44.0°, and 52.4° by the Multispectral Visible Imaging Camera (MVIC) on the New Horizons spacecraft during 2023, 2010, and 2019, respectively. New Horizons imaged Uranus at a distance of about 24--70\,AU (2023) in four color filters, with bandpasses of 400-550~nm, 540-700~nm, 780-975~nm, and 860-910~nm. High-phase-angle observations are of interest for studying the energy balance of Uranus, constraining the atmospheric scattering behavior, and understanding the planet as an analog for ice giant exoplanets. The new observations from New Horizons provide access to a wider wavelength range and different season compared to previous observations from both Voyager spacecraft. We performed aperture photometry on the New Horizons observations of Uranus to obtain its brightness in each photometric band. The photometry suggests that Uranus may be darker than predicted by a Lambertian phase curve in the Blue and Red filters. Comparison to simultaneous low-phase Hubble WFC3 and ground-based community-led observations indicates a lack of large-scale features at full-phase that would introduce variation in the rotational light curve. The New Horizons reflectance in the Blue (492~nm) and Red (624~nm) filters does not exhibit statistically significant variation and is consistent with the expected error bars. These results place new constraints on the atmospheric model of Uranus and its reflectivity. The observations are analogous to those from future exoplanet direct-imaging missions, which will capture unresolved images of exoplanets at partial phases. These results will serve as a ``ground-truth" with which to interpret exo-ice giant data.

\end{abstract}


\section{Introduction} \label{sec:intro}

    The only measurements of heat balance in Uranus and Neptune came from Voyager 2 flyby data in 1986 and 1989, respectively. 
    During these observations, Uranus was experiencing its southern summer solstice and Neptune was beginning to approach northern hemisphere winter solstice \citep{Meeus1997JBAA..107..332M}. 
    The Voyager observations ultimately left major questions about why each of the planets is so different despite their similar bulk composition and size.
    For example, Uranus appears to have little internal heat, while Neptune has a substantial internal reservoir \citep{Pearl1990-od, Pearl1991-xv}. 
    To accurately calculate the energy balance of the planets, both the reflected sunlight and the thermal emission need to be measured  at multiple epochs, over many phase angles, and with broad wavelength coverage. 

    For the reflected light component, Voyager 2’s best observations came from the IRIS instrument’s visible/NIR channel. 
    IRIS visible/NIR measurements of Uranus’ full disk were from phase angles around 20\degr{} and 150\degr{}.
    IRIS measurements of selected regions on Uranus were also taken at three intermediate phase angles around 31\degr{}, 54\degr{}, and 108\degr{} \citep{Pearl1990-od}. 
    The IRIS instrument's radiometer and interferometer were well-calibrated to an internal target plate of known scattering properties and a blackbody function, respectively \citep{Hanel1980ApOpt..19.1391H, Hanel1981JGR....86.8705H, Pearl1990-od}.
    These data were supplemented by Voyager 1 imaging observations through the narrow angle camera in multiple visible bands, acquired from large distances throughout the 1980s, at 13 phase angles between 25\degr{} and 107\degr{} \citep{wenkert2022AGUFM.P32E1865W, wenkert2023PDS}. 
    However, the Voyager 1 imaging results suffered from low SNR and a radiometric calibration that is less precisely known than those of more modern imaging systems, and were limited to wavelengths shortward of about 600\,nm.
    
    As part of New Horizons' second extended mission, the spacecraft was uniquely positioned to study Uranus and Neptune at high phase angles, which are not accessible from any other ground-based or space-based facility. 
    A series of Uranus observations was conducted in September 2023 by New Horizons, which captured the planet at a phase angle of 43.9°. Simultaneous full-phase observations of Uranus were captured by the Hubble Space Telescope (see Section~\ref{sec:hubble_imaging}). The New Horizons Science Team also invited the ground-based amateur astronomy community to support simultaneous observations and received submissions from community observers around the globe (see Section~\ref{sec:community_obs}). Two additional high-phase-angle scans were captured by New Horizons in 2010 and 2019, with phase angles of 44.0° and 53.4°, respectively.
    
    Here we build on the existing literature of Uranus observations from the Voyager missions (e.g., \citealt{smith1986Sci...233...43S, Pearl1990-od}) with new data to refine our understanding of the planet's energy budget and heat balance, better constrain its atmospheric properties, and provide additional constraints for future exoplanet studies. We note that these observations are not a direct one-to-one comparison with the Voyager flyby data as they were taken at a different season for Uranus (currently approaching its southern hemisphere winter solstice), and many temporal variations could have occurred in between. 

    In Section \ref{sec:methods}, we introduce the New Horizons Ralph Instrument and Multispectral Visible Imaging Camera \citep{Reuter2008SSRv..140..129R}, the data reduction and analysis method that we used to gather photometry of Uranus, and the supporting observations taken with Hubble and by community observers. 
    In Section \ref{sec:results}, we present the measured photometry from New Horizons and a comparison to the spectrum of Uranus. We discuss our results and implications both for our Solar System and for an exoplanet context in Section \ref{sec:discussion} before concluding in Section \ref{sec:conclusion}.

\section{Methods} \label{sec:methods}

\begin{figure*}
    \centering
    \includegraphics[width=0.85\textwidth]{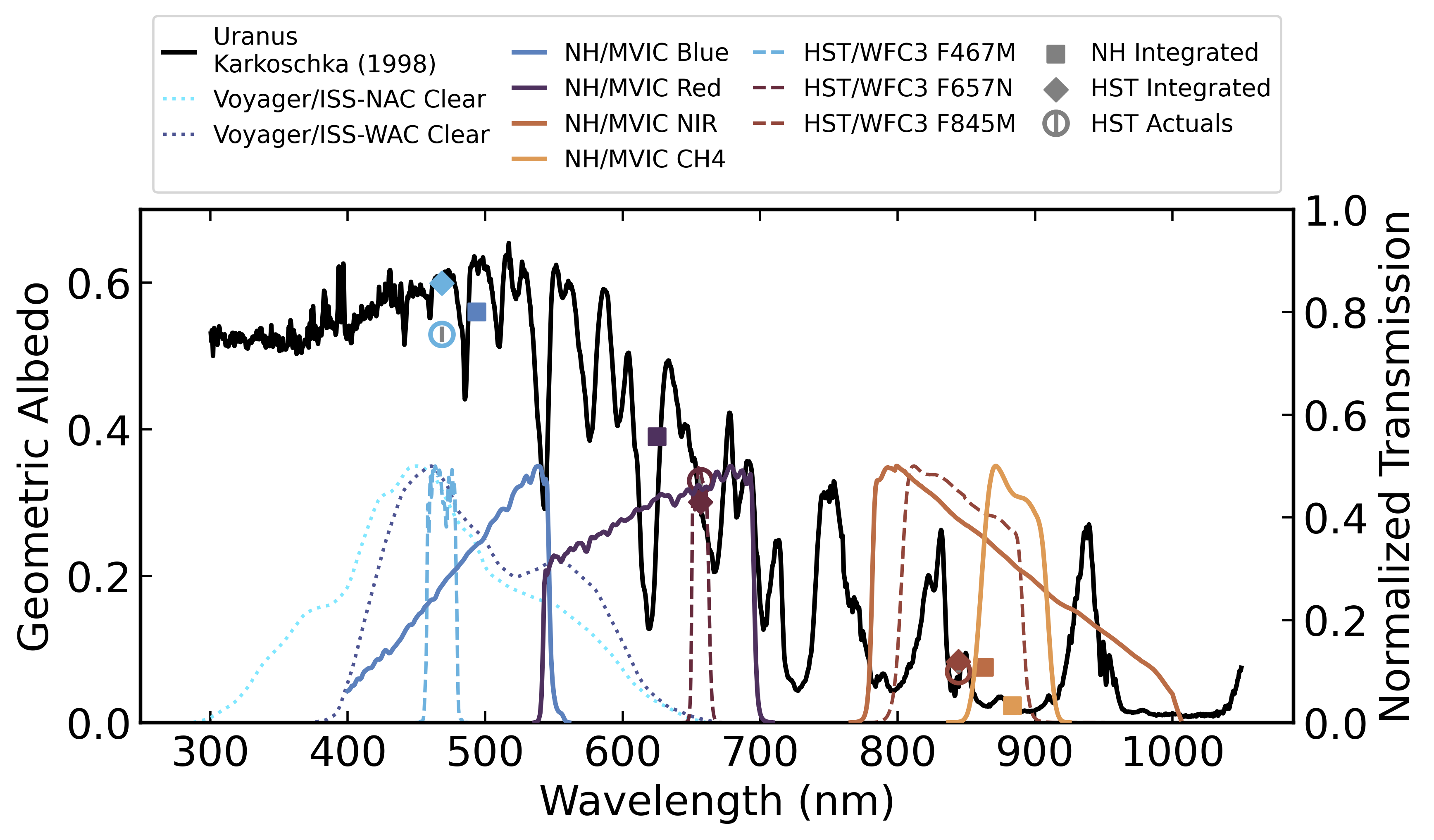}
    \caption{Geometric albedo of Uranus compared to select filter transmission curves of Voyager, New Horizons and HST. Filter transmission curves have been normalized to a max value of 0.5 for display purposes. The expected filter-integrated values for the New Horizons and HST observations are shown with squares and diamonds, while the actual observations made with the HST campaign are shown in circles with errorbars (see Section~\ref{sec:hubble_imaging}). New Horizons data add much needed spectral coverage at critical phase angles.}
    \label{fig:spectrum}
\end{figure*}

\subsection{Observation Platform}

    The Ralph instrument is mounted on the exterior of the New Horizons spacecraft \citep{Reuter2008SSRv..140..129R}. 
    It consists of a single 75-\,mm $f$/8.7 off-axis reflecting telescope that feeds both the Multispectral Visible Imaging Camera (MVIC) and the Linear Etalon Imaging Spectral Array (LEISA), with the light for MVIC being reflected by a dichroic beamsplitter.  
    
    MVIC is a visible to near-IR (NIR) broadband imager, with four color filters designated as Blue (400--550\,nm),  Red (540--700\,nm), near-IR (780--975\,nm) and methane (860--910\,nm). The filters are shown in Figure~\ref{fig:spectrum} against the \citet{karkoschka1998Icar..133..134K} albedo spectrum of Uranus and other notable filters.  
    Each filter is affixed to a 5024$\times$32 pixel CCD, providing a static FOV of 5\dotdeg{}7$\times$0\dotdeg{}037, with an instantaneous FOV of 20 microradians.  These CCDs are installed adjacent to one another in the MVIC focal plane and operated in time delay integration mode (TDI).  There is no scan mirror or instrument pointing platform; the entire spacecraft is slewed so as to move a target scene across the short axis of the CCD arrays at a rate corresponding to that at which accumulating charges are transferred from one pixel to the next toward the readouts.  Each such scan produces four separate color images that must subsequently be registered together to produce a four-color image cube.

\subsection{Data and Reduction} \label{subsec:data}

    New Horizons obtained six scans of Uranus with the four MVIC color filters on 2023 September 16, one scan on 2019 September 2, and one scan on 2010 June 23.
    The observation details are listed in Table~\ref{tab:observations}. 
    Each of the 2023 scans was taken at a distance from Uranus of approximately 69.5\,AU, with a total exposure time of 0.765\,s (summed over the 32 TDI rows).
    The 2019 scans were taken at a distance from Uranus of $\sim$54.5\,AU with an exposure time of 0.768\,s. The 2010 scans were taken at a distance of $\sim$23.3\,AU with an exposure time of 1.70\,s.
    All MVIC observations used here are Level 2b data pipeline products, which have been bias and flat-field corrected following the standard New Horizons data reduction and calibration routine \citep{Peterson-2008-NHSOC, howett2017Icar..287..140H}.  
    
    Due to instrument pointing and small discrepancies between the spacecraft scan rate and CCD readout rate, the predicted location of Uranus is not always the precise location in the image. The Right Ascension (RA) and Declination (Dec) of Uranus at the time of imaging was computed using SPICE routines. These were then compared to the RA/Dec of known stars across the image, to determine where Uranus' RA/Dec was in the image. 
    
    We determine the final pixel location of Uranus in the MVIC images before performing aperture photometry by using background stars to triangulate the positions between sets of images and across each of the filters. After identifying nearby stars to Uranus' position in each image, we compute the offset of each image from a reference image and apply the same offset to the location of Uranus. This allows us to approximate Uranus' location even in filters where Uranus is not visible above the background noise. 
    
\begin{figure*}
    \centering
    \includegraphics[width=0.9\textwidth]{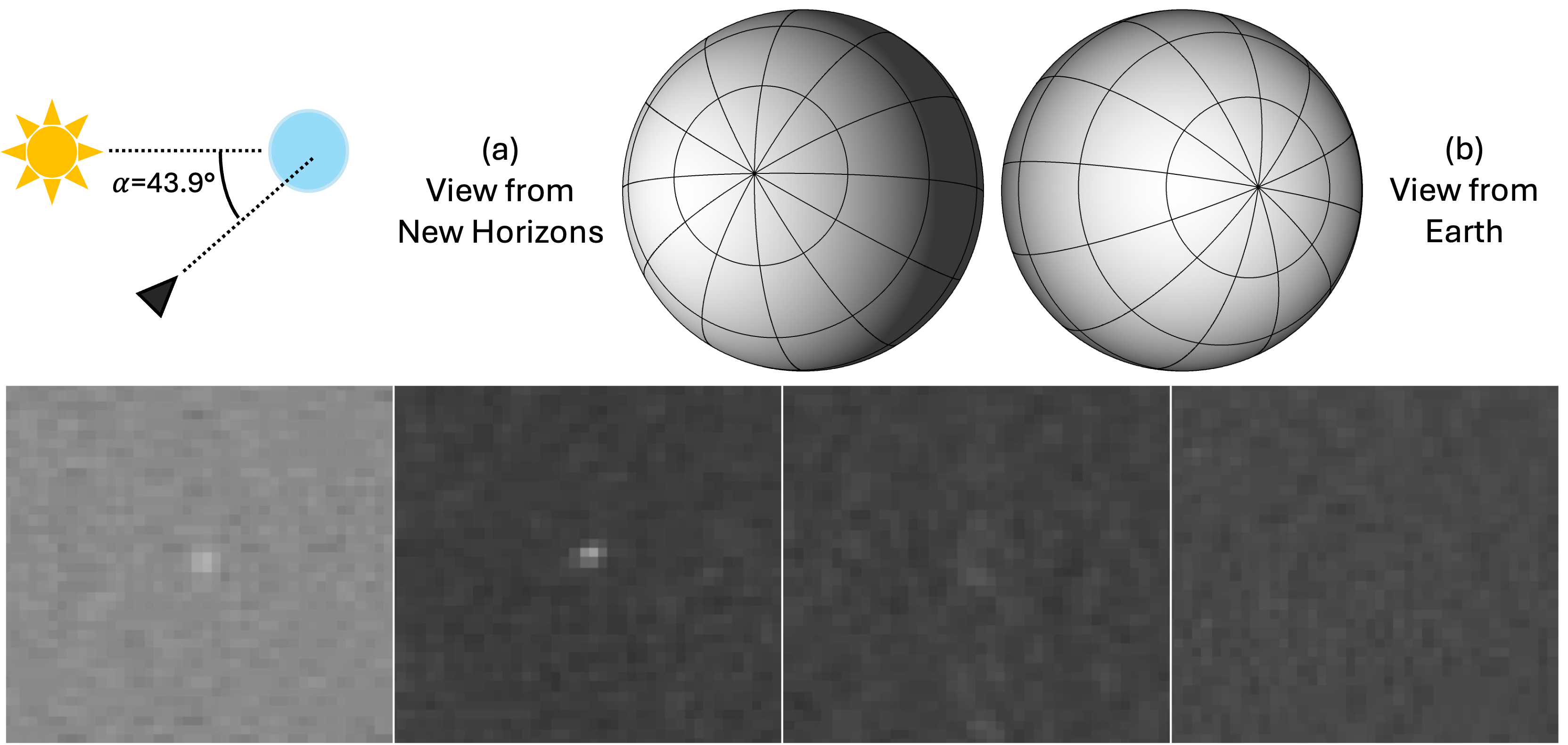}
    \caption{Top: Observing geometry of Uranus in 2023 from (a) New Horizons and (b) from Earth where celestial north is up. A diagram of the phase angle ($\alpha$) geometry between New Horizons, Uranus, and the Sun is shown in the top left
    Bottom: Destriped MVIC frames of Uranus from scan 0557172410 on 16 September 2023. From left to right, each panel is centered on detections (or non-detections) of Uranus in the Blue,  Red, NIR, and CH$_{4}$ filters. The data are presented in the native pixel orientation.}
    \label{fig:NH_uranus}
\end{figure*}

    Additional corrections were required of the Level 2b data before performing aperture photometry. 
    The data contained notable striping artifacts that required removal prior to our analysis. 
    We destriped the data by performing median subtraction across the rows of the MVIC windowed observation. Sub-panels of fully corrected images across each of the filters are shown in Figure~\ref{fig:NH_uranus}, centered on Uranus. The data show strong detections of Uranus in the Blue and  Red filters, weak detections in the NIR filter, and non-detections in the methane band. 

\subsection{Photometry} \label{subsec:photometry}

    After destriping, we determined the minimum aperture size required for photometry by testing a range of apertures with radii between 1 to 8 pixels in 0.5-pixel increments. 
    We generated curves to measure the flux of several point sources, including Uranus, across the aperture ranges to determine the required cutoff in aperture size. 
    We tested aperture sizes across each of the filters to determine the appropriate aperture per wavelength band of observation. 
    Aperture radii were selected to be the minimum aperture radius where the plateau in flux began. 
    Our tests resulted in aperture radii of 2.5, 3, 4, and 4 pixels for the Blue,  Red, NIR, and CH$_4$ filters, respectively. 
    We used inner annulus radii of 5, 6, 8, and 8 pixels and outer annulus radii of 18 pixels for each filter. 
    We use the \texttt{photutils.aperture\_photometry} routine to perform aperture photometry of Uranus at the DN level \citep{bradley_2023}.
    
    We assume that our only sources of noise are Poisson noise and read noise. We use the standard equation for noise computation assuming that the sky background and the read noise are the only other sources \citep[per][dark current is a negligible noise source]{Reuter2008SSRv..140..129R}. 
    We convert image counts to electrons using the gain (58.6\,e-/DN, \citealt{Reuter2008SSRv..140..129R}) and conducted aperture photometry in electron space, thus allowing us to define the error on each observation as the sum of the square root of the number of electrons/pixel of the object and sky annulus and the read noise (30\,electrons, \citealt{Reuter2008SSRv..140..129R}). Thus,
    \begin{equation}
        err = \sqrt{S_{signal} + (S_{Sky,pix} + n_r^2) A_{ap} )}
    \end{equation}
    where $S_{signal}$ is the object flux in electrons, $S_{Sky,pix}$ is the per pixel flux in the sky annulus, $n_r$ is the read noise, and $A_{ap}$ is the area of the aperture in pixels$^2$. 
    
    As a result of the New Horizons photometric pipeline there are occasionally negative values in the images. To proceed with standard noise estimation we have added the minimum value from the image to the entire image array to remove any negative values in the region surrounding our object.
    Typical errors on our values from the 2023 data are on the order of 18\% to 32\% in the Red and Blue filters, respectively, where Uranus is brighter, 313\% in the NIR filter, and 318\% in the CH$_4$ filter, where MVIC receives little to no flux from Uranus.
    
    To convert the MVIC measurements of Uranus from number of counts (DN) to the radiance factor (I/F) we first determined the spatial scale ($d$) of the observations by multiplying the target-observer distance in kilometers ($D$) (see Table~\ref{tab:observations}) by the pixel scale of MVIC in millimeters ($P$) divided by the focal length in millimeters ($L$):
    
    \begin{equation}
        d = D * \frac{P}{L}
    \end{equation}
    
    \noindent We calculated the radius of Uranus in pixels by dividing the radius of the target ($R_T$) by the spatial scale:
    
    \begin{equation}
        R_{pix} = \frac{R_T}{d} 
    \end{equation}
    
    \noindent In this case, $R_T$ is equal to the radius of Uranus. 
    We take $R_T$ to be 25,362\,km, which is the mean radius of Uranus given by the JPL Horizons database\footnote{\url{https://ssd.jpl.nasa.gov/planets/phys\_par.html}} \citep{archinal2018CeMDA.130...22A}. 
    The area of Uranus in pixels is given by:
    
    \begin{equation}\label{eqn:Apix}
        A_{pix} = \pi R_{pix}^2
    \end{equation}
    
    To compute the radiance, we first calculated the flux of the target ($F_\mathrm{fil}$) from the count rate of the observation and a calibration-dependent keyword: 
    
    \begin{equation}
        F_\mathrm{fil} = \frac{c}{R_\mathrm{U,fil}} 
    \end{equation}
    where $c$ represents the counts obtained from aperture photometry divided by the exposure time of the observation and has units of DN s$^{-1}$, and $R_\mathrm{U,fil}$ is the MVIC calibration-dependent keyword. 
    The keyword was calculated using the spectrum of Uranus from \citet{karkoschka1998Icar..133..134K}. 
    We used the extended source keyword calculated for Uranus, which has units of DN s$^{-1}$ / (erg s$^{-1}$ cm$^{-2}$ \AA $^{-1}$ sr$^{-1}$).
    Further detail regarding the MVIC keyword can be found in \cite{howett2017Icar..287..140H}. 
    
    To account for the full solid angle subtended by the target, we divided the resulting flux, $F_\mathrm{fil}$, by the target area in pixels from Equation~\ref{eqn:Apix}:
    
    \begin{equation}
        I_\mathrm{fil} = \frac{F_\mathrm{fil}}{A_{pix}}
    \end{equation}
    Finally, we computed the I/F in each filter using the following equation:
    \begin{equation}\label{eqn:I/F}
        \frac{I}{F} = \frac{\pi I_\mathrm{fil} r^2}{F_{\sun}}
    \end{equation}
    
    \noindent where $r$ is the target's heliocentric distance in AU and $F_{\sun}$ is the solar flux at 1\,AU in the filter of the observation, calculated by convolving the solar spectrum with the MVIC filter transmission curve. The solar reference spectra used here is described in \cite{Colina1996AJ....112..307C}. 

\subsection{Hubble Support Imaging}\label{sec:hubble_imaging}

\begin{figure*}
    \centering
    \includegraphics[width=0.85\textwidth]{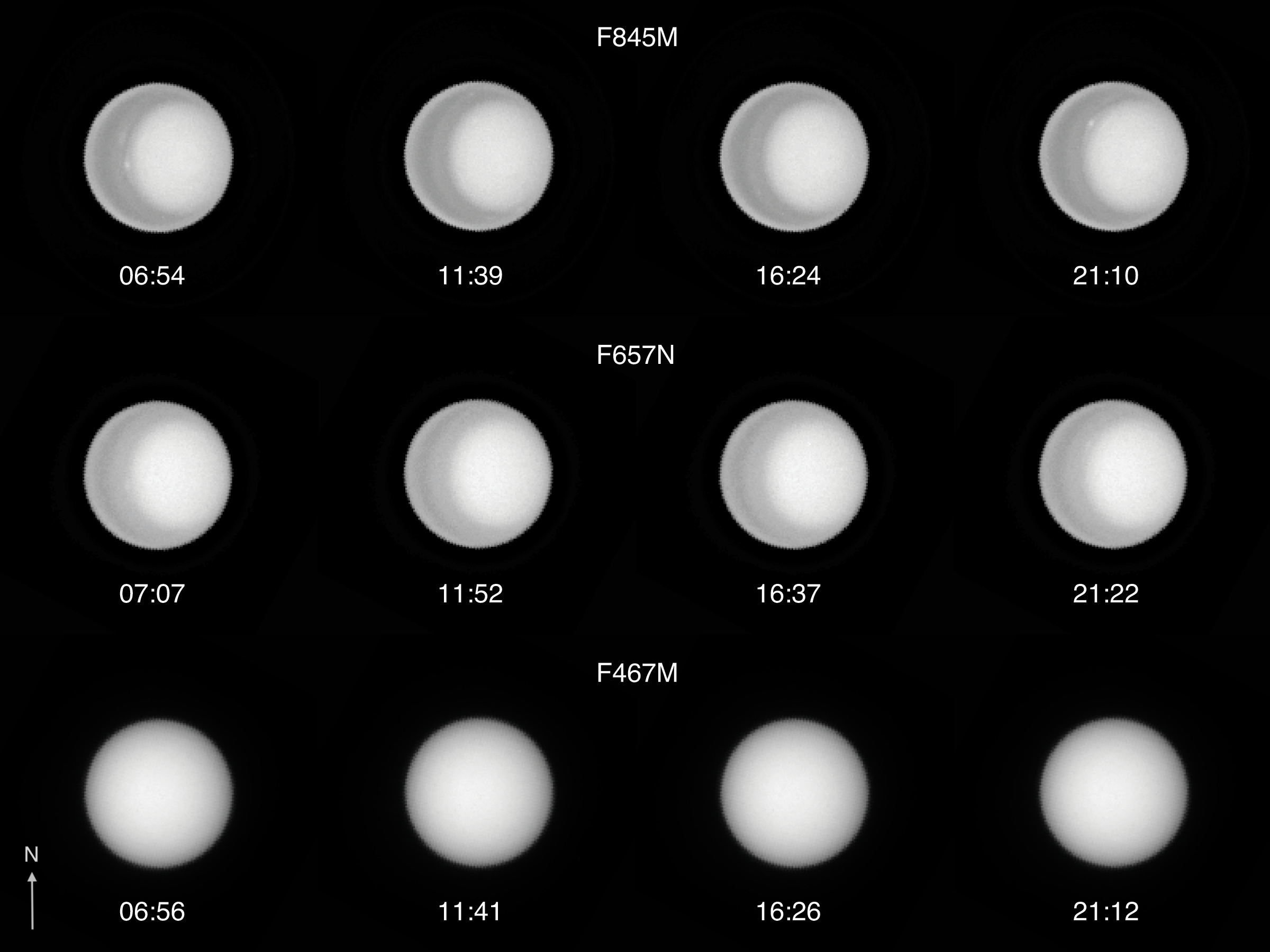}
    \caption{Hubble images from 17 September 2023 in the F845M, F657N, and F467M filters, taken as part of program GO17294.  The approximate image UTC start times are listed for each frame (rounded down to the nearest minute), and the F845M and F657N images have an unsharp mask applied to enhance cloud features. Hubble images have been rotated from their native pixel orientation so that celestial north is oriented up (as indicated by the arrow in the bottom left corner).}
    \label{fig:Hubble}
\end{figure*}

Hubble observations of Uranus were also obtained between 2023 September 17 06:54 UTC and 2023 September 18 21:34 UTC as part of the Outer Planet Atmospheres Legacy (OPAL) program \citep{Simon2015ApJ...812...55S}. 
These observations covered two full Uranus rotations in seven filters (from $\sim$460 to 870\,nm) and provide low-phase-angle light curves and context imaging for the New Horizons observations. 
The Hubble Wide Field Camera 3 (WFC3) filters used by OPAL are narrow-and medium-band passes, and none is a perfect match to the MVIC filters.  
The obtained WFC3 filters that best overlap those of the MVIC observations are F467M ($\sim$458-478\,nm), F657N ($\sim$651-663\,nm), and F845M ($\sim$804-884\,nm); although Hubble methane band observations were obtained, they used 619-\,nm and 727-\,nm filters, rather than the 889-\,nm band used by MVIC. 

Hubble images are processed using the standard calibration pipeline, post-pipeline cosmic ray cleaning, and defringing  as described in \cite{Simon2015ApJ...812...55S}. 
The images are navigated using an iterative limb-fitting procedure and converted to I/F using the photometric conversion values in each image header, the solar flux in the bandpass, and the planet's heliocentric range. 
Flux calibration is accurate to about 2\%, but photometric stability is $\sim$0.5\% \citep{Calamida2022AJ....164...32C}.  Images are then mapped to create publicly accessible global maps (\url{https://archive.stsci.edu/hlsp/opal}; \citealt{HST_OPAL_DOI}). 
Light curves can be obtained from individual full-disk views, albeit with coarse time coverage.  

In the September 2023 dataset, Uranus' approximate disk-averaged I/F was 0.53, 0.33, and 0.07 for the F467M, F657N, and F845M filters, respectively, with a 2\% uncertainty (see Figure~\ref{fig:spectrum}). 
The disk is featureless at Blue wavelengths, with the polar cap becoming more prominent at  Red and infrared wavelengths, as shown in Figure~\ref{fig:Hubble}. 
A small bright feature is seen near the polar haze cap, particularly visible in the F845M filter. However the disk-averaged brightness showed no variations over two Uranus rotations, in any filters, above the photometric precision of $\sim$0.5\%.
Thus, any New Horizons measurements should not be affected by which Uranus longitudes were visible.

\subsection{Community Observations}\label{sec:community_obs}

\begin{figure*}
    \centering
    \includegraphics[width=0.85\textwidth]{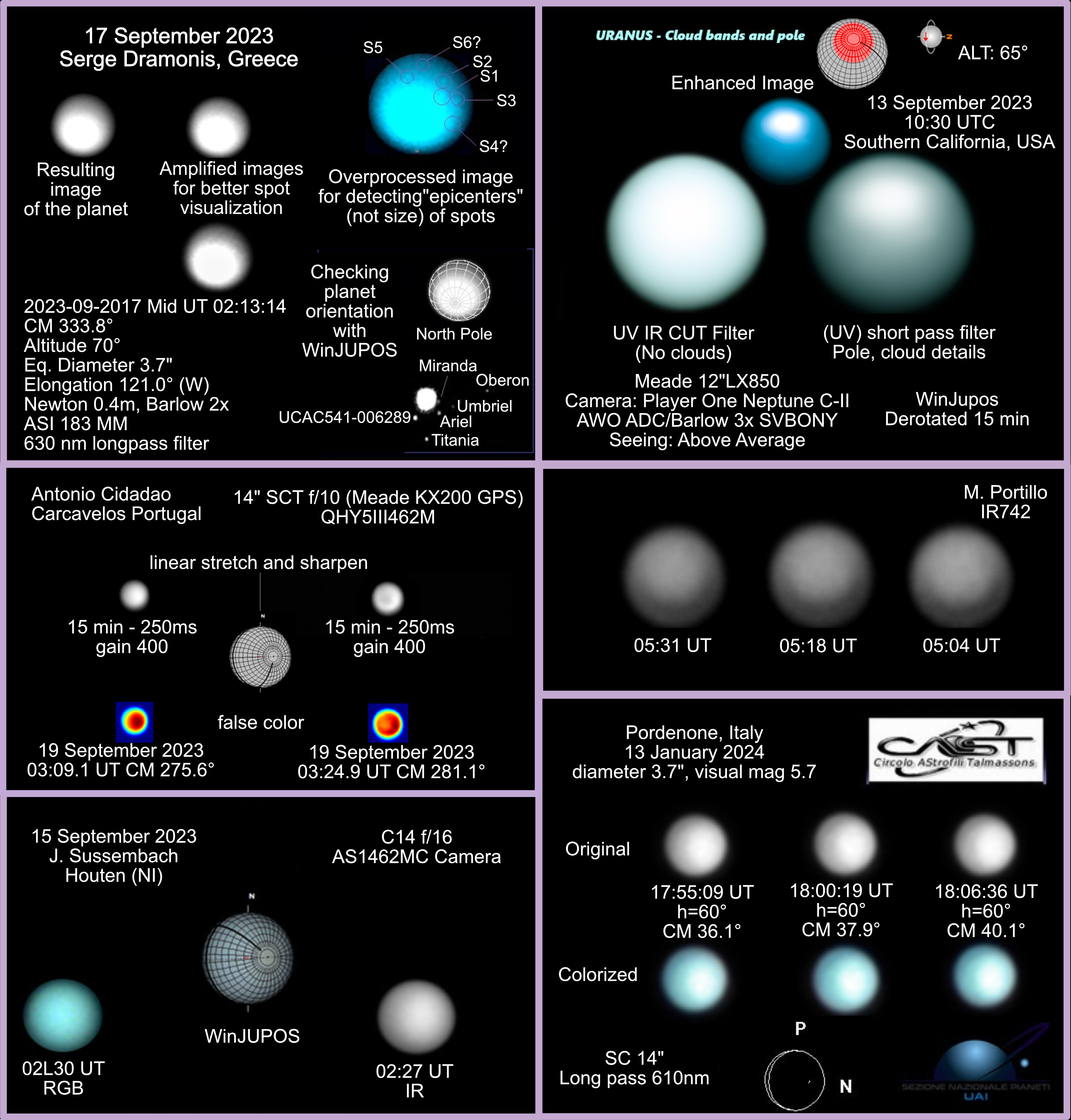}
    \caption{Sample 
    image submissions from the ground-based community observer campaign. Some images were processed more than others and some were submitted as series of nights. All images are presented in their native pixel orientation as submitted by observers. Observations were collected in both the optical and near-IR from 12-16 inch diameter telescopes around the globe. A list of individuals who submitted images for this campaign and included their names in their submissions can be found in the acknowledgements section. Some images were received without being able to identify the observer, though observational details were provided. We are very grateful all who contributed.}
    \label{fig:publicImages}
\end{figure*}

In an effort to engage the general public in the mission and these observations, we organized a ground-based campaign around the time of the New Horizons and Hubble observations. 
We requested individuals to observe Uranus and Neptune prior to, during, and after the spacecraft observations, and to post the observations to Facebook, X, or to upload them to a repository we hosted at filerequestpro.com.
We received $\sim$100 image submissions (some as image series) from around the world including from groups in Austria, Australia, Canada, Greece, Italy, Portugal, and various sites in the United States. 
Images included data from mid-July through the end of 2023 in optical and infrared wavelengths with telescopes ranging from 12 to 16 inches in diameter. 
Figure \ref{fig:publicImages} shows sample images. 
The main goal of the campaign was to look for hints of longer-term cloud patterns on these planets which might aid in providing context for interpreting the spacecraft datasets. Some of the community observations were able to resolve the disk of Uranus and capture the polar structure, which can provide a glimpse into the atmospheric dynamics over time.

\section{Results} \label{sec:results}

\begin{deluxetable*}{c|c|ccccccr}
\tabletypesize{\footnotesize}
\tablecaption{Table of Observations\label{tab:observations}}
\tablehead{
    \colhead{Scan} & \colhead{Obs.\,Time} & \colhead{Distance} & \colhead{Exp.\,Time} & \colhead{Sub-Observer} & \colhead{Sub-Observer}& \colhead{Phase Angle} & \colhead{I/F} & \colhead{Filter} \\ 
    \colhead{} & \colhead{} & \colhead{(km)} & \colhead{\textbf{(s)}} & \colhead{Latitude (°)} & \colhead{Longitude (°)} & \colhead{(°)} & \colhead{} & \colhead{} 
    }
    
\startdata
\multirow{4}{*}{139565811} & \multirow{4}{*}{\parbox{1.8cm}{2010 Jun 23 02:25:16.627}} & \multirow{4}{*}{3.62E09} & \multirow{4}{*}{1.70} & \multirow{4}{*}{54.77} & \multirow{4}{*}{147.38} & \multirow{4}{*}{44.0} & 0.366$\pm$0.008 & Blue \\
 &  &  &  &  &  &  & 0.100$\pm$0.002 &  Red \\
 &  &  &  &  &  &  & 0.010$\pm$0.001 & NIR \\
 &  &  &  &  &  &  & $<$0.018 & CH4 \\
 \hline
\multirow{4}{*}{429771832} & \multirow{4}{*}{\parbox{1.8cm}{2019 Sep 02 23:12:07.456}} & \multirow{4}{*}{8.16E09} & \multirow{4}{*}{0.768} & \multirow{4}{*}{80.14} & \multirow{4}{*}{167.16} & \multirow{4}{*}{52.4} & 0.296$\pm$0.075 & Blue \\
 &  &  &  &  &  &  & 0.103$\pm$0.013 &  Red \\
 &  &  &  &  &  &  & $<$0.014 & NIR \\
 &  &  &  &  &  &  & $<$0.185 & CH4 \\
 \hline
\multirow{4}{*}{0557172410} & \multirow{4}{*}{\parbox{1.8cm}{2023 Sep 16 12:15:06.888}} & \multirow{4}{*}{1.04E10} & \multirow{4}{*}{0.765} & \multirow{4}{*}{73.39} & \multirow{4}{*}{41.5} & \multirow{4}{*}{43.9} & 0.413$\pm$0.111 & Blue \\
 &  &  &  &  &  &  & 0.120$\pm$0.021 &  Red \\
 &  &  &  &  &  &  & 0.033$\pm$0.019 & NIR \\
 &  &  &  &  &  &  & $<$0.132 & CH4 \\
 \hline
\multirow{4}{*}{0557182706} & \multirow{4}{*}{\parbox{1.8cm}{2023 Sep 16 15:06:42.888}} & \multirow{4}{*}{1.04E10} & \multirow{4}{*}{0.765} & \multirow{4}{*}{73.39} & \multirow{4}{*}{101.22} & \multirow{4}{*}{43.9} & 0.315$\pm$0.105 & Blue \\
 &  &  &  &  &  &  & 0.114$\pm$0.025 &  Red \\
 &  &  &  &  &  &  & 0.013$\pm$0.019 & NIR \\
 &  &  &  &  &  &  & $<$0.108 & CH4 \\
 \hline
\multirow{4}{*}{0557193002} & \multirow{4}{*}{\parbox{1.8cm}{2023 Sep 16 17:58:18.888}} & \multirow{4}{*}{1.04E10} & \multirow{4}{*}{0.765} & \multirow{4}{*}{73.39} & \multirow{4}{*}{160.94} & \multirow{4}{*}{43.9} & 0.374$\pm$0.114 & Blue \\
 &  &  &  &  &  &  & 0.129$\pm$0.022 &  Red \\
 &  &  &  &  &  &  & $<$0.018 & NIR \\
 &  &  &  &  &  &  & $<$0.0.145 & CH4 \\
 \hline
\multirow{4}{*}{0557203298} & \multirow{4}{*}{\parbox{1.8cm}{2023 Sep 16 20:49:54.888}} & \multirow{4}{*}{1.04E10} & \multirow{4}{*}{0.765} & \multirow{4}{*}{73.39} & \multirow{4}{*}{220.66} & \multirow{4}{*}{43.9} & 0.318$\pm$0.104 & Blue \\
 &  &  &  &  &  &  & 0.132$\pm$0.020 &  Red \\
 &  &  &  &  &  &  & $<$0.016 & NIR \\
 &  &  &  &  &  &  & $<$0.138 & CH4 \\
\hline
\multirow{4}{*}{0557213594} & \multirow{4}{*}{\parbox{1.8cm}{2023 Sep 16 23:41:30.888}} &  \multirow{4}{*}{1.04E10} & \multirow{4}{*}{0.765} & \multirow{4}{*}{73.39} & \multirow{4}{*}{280.38} & \multirow{4}{*}{43.9} & 0.299$\pm$0.113 & Blue \\
 &  &  &  &  &  &  & 0.099$\pm$0.026 &  Red \\
 &  &  &  &  &  &  & $<$0.020 & NIR \\
 &  &  &  &  &  &  & $<$0.121 & CH4 \\
\hline
\multirow{4}{*}{0557223890} & \multirow{4}{*}{\parbox{1.8cm}{2023 Sep 17 02:33:06.888}} &  \multirow{4}{*}{1.04E10} & \multirow{4}{*}{0.765} & \multirow{4}{*}{73.39} & \multirow{4}{*}{340.09} & \multirow{4}{*}{43.9} & 0.380$\pm$0.120 & Blue \\
 &  &  &  &  &  &  & 0.125$\pm$0.020 &  Red \\
 &  &  &  &  &  &  & 0.017$\pm$0.023 & NIR \\
 &  &  &  &  &  &  & $<$0.118 & CH4 \\
\hline
\enddata
\tablecomments{In the case of non-detections (such as in the NIR and CH$_4$ filters) we have given the one-sigma upper limit on the I/F value.}
\end{deluxetable*}

\begin{figure*}
    \centering
    \includegraphics[width=0.74\textwidth]{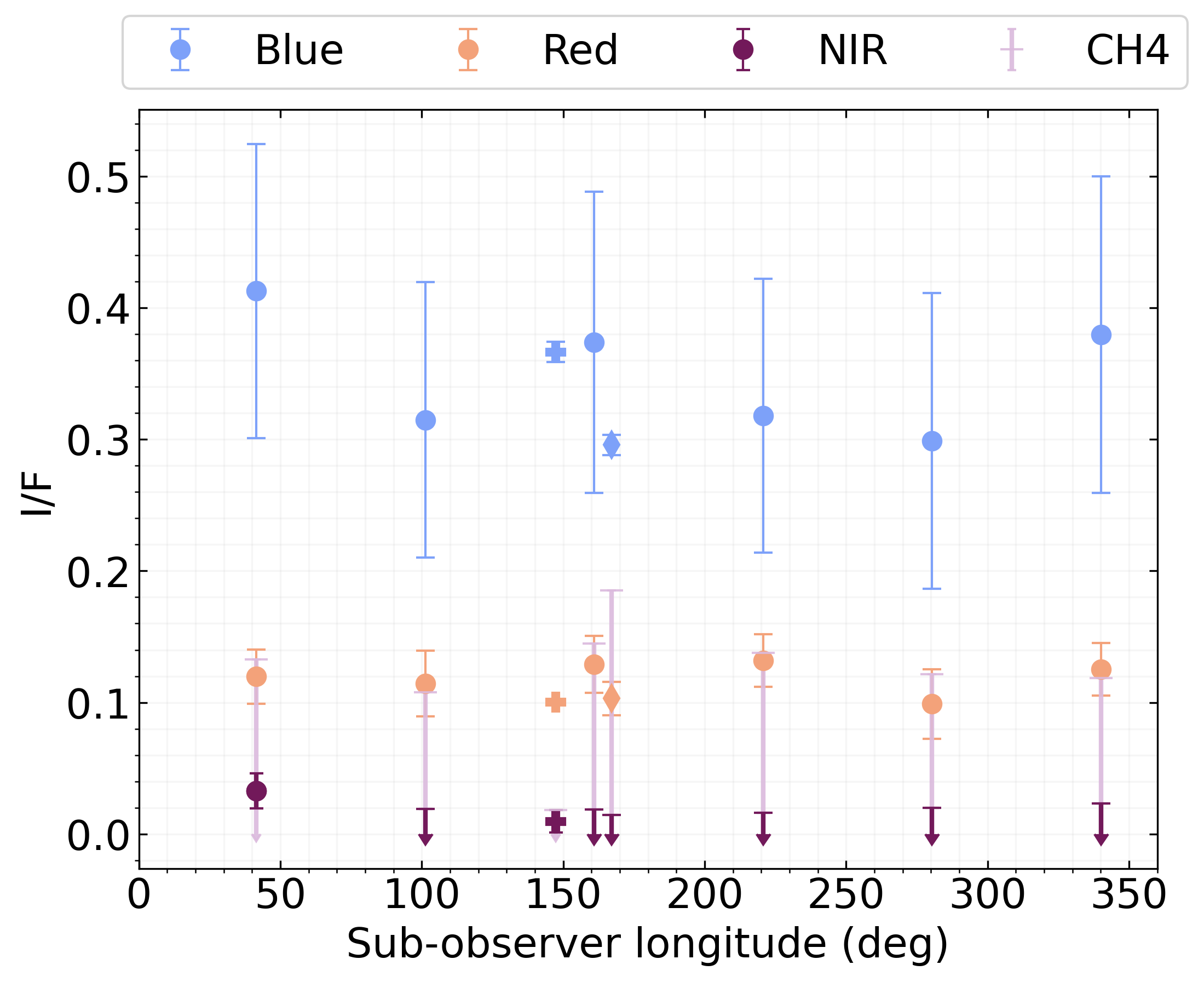}
    \caption{I/F as function of sub-observer longitude for each of the MVIC filters. The 2023 data are indicated by circles, the 2019 data by diamonds, and the 2010 data by plus signs. Error bars are one-sigma. In the case of non-detections (such as in the NIR and CH$_4$ filters) we show a one-sigma upper limit. I/F values are also listed in Table~\ref{tab:observations}, along with observation details.}
    \label{fig:I/F_SubObs}
\end{figure*}

\begin{figure*}
    \centering
    \includegraphics[width=0.9\textwidth]{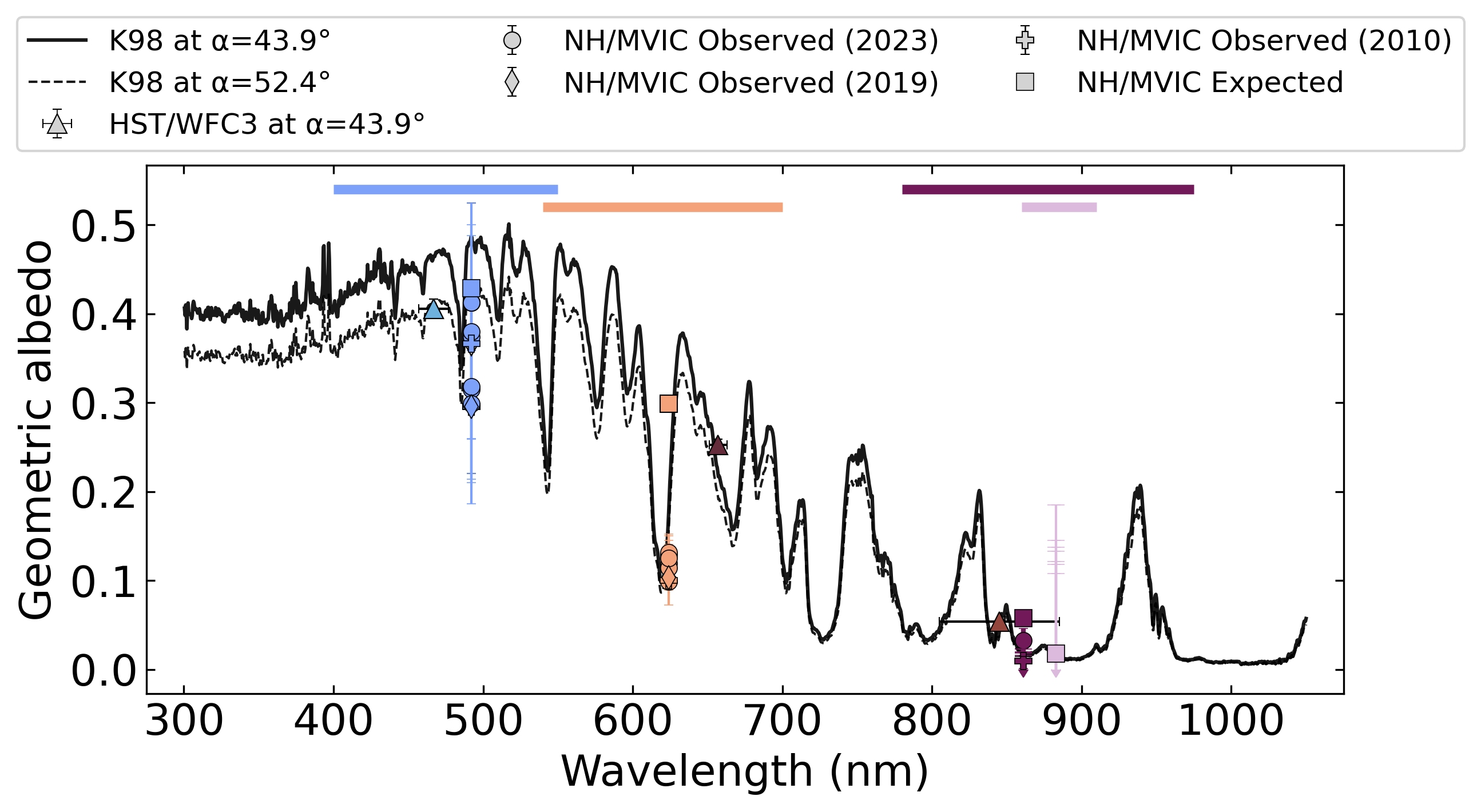}
    \caption{Expected and observed NH/MVIC I/F compared to the \cite{karkoschka1998Icar..133..134K} (K98) geometric albedo spectrum scaled to a phase of $\alpha=43.9\degr{}$ and $\alpha=53.4\degr{}$ with a Lambertian phase function. Squares indicate the expected NH/MVIC values, calculated by convolving the $43.9\degr{}$ albedo spectrum with the MVIC filters. NH/MVIC observations are plotted with circles (for the 2023 scans), diamonds (for the 2019 scan), and plus signs (for the 2010 scan). Horizontal lines indicate the bandwidth of the Blue,  Red, NIR, and CH$_4$ filters. The disk-averaged I/F data from the HST/WFC3 support imaging have also been scaled to a phase of $43.9\degr{}$ with a Lambertian phase function and are shown here with triangles for comparison.}
    \label{fig:IF_Karkoschka}
\end{figure*}

We show the resulting I/F obtained in each filter and scan in Table~\ref{tab:observations}. 
The flux of Uranus measured by New Horizons exhibits minor variation as a function of the sub-observer longitude. 
The largest fluctuations are exhibited in the Blue filter, where Uranus is the brightest. 
Figure~\ref{fig:I/F_SubObs} shows the I/F values for each of the six scans and across all four MVIC filters as a function of longitudes. 
Longitudes and latitudes are taken from JPL Horizons\footnote{\url{https://ssd.jpl.nasa.gov/horizons/}}. 
The 2023 observations were taken over an approximately 14-hour period, which spans nearly one full rotation of the planet (17.2 hours). 
Fluctuations are significantly less in the  Red, NIR, and CH$_4$ filters, where the signal from Uranus also decreases. The 2010 and 2019 observations provide two additional measurements near the $\sim$150$\degr{}$ sub-observer longitude range. These measurements are consistent with the 2023 observations and do not suggest significant fluctuation between 2010 to 2023.

The MVIC observations are plotted against the spectrum of Uranus in Figure~\ref{fig:IF_Karkoschka}. The spectrum of Uranus at a phase angle of 43.9\degr{} is shown in black.
We approximated the geometric albedo spectrum of Uranus at 43.9\degr{} phase angle by applying a Lambertian phase function to the full phase spectrum. The full-phase spectrum was measured empirically by \cite{karkoschka1998Icar..133..134K}. We have also plotted the albedo spectrum approximated at 52.4\degr{} by a Lambertian phase function with a black dashed line.
MVIC photometric points are shown in comparison to simulated photometry calculated using the \cite{karkoschka1998Icar..133..134K} spectrum convolved with MVIC's filters. 
The approximate disk-averaged I/F from the Hubble support imaging is also plotted for comparison.

\section{Discussion} \label{sec:discussion}

While the Hubble observations and amateur observations of Uranus show the polar cap of the planet with few features, the disk-integrated brightness of these images suggest no statistical variability in Uranus' flux over the course of a rotation at essentially a phase angle of zero. 
We can examine Figure~\ref{fig:I/F_SubObs} for statistically significant longitudinal variations in I/F.
We fit the Blue data points with a best-fit flat line (I/F $\approx$ 0.33) and find that all data points are within 1$\sigma$. 
The  Red data points all lie within approximately 2$\sigma$ of the best-fit flat line (I/F $\approx$ 0.11). 
Additional rotations may have allowed for a more robust detection of any variation if said variation was present, such as the bright feature seen near the polar haze cap in the F845M filter. 
This feature is at approximately 64\degr{} E longitude and may have been captured in our first 2023 MVIC scan.

As seen in Figure~\ref{fig:IF_Karkoschka}, the data suggest that Uranus at high-phase-angle is already darker than predicted by a Lambertian phase curve, particularly in the Blue and Red filters where we have good signal.
The signal of Uranus in the NIR and CH$_4$ filters is too small to say anything of statistical significance.

Previous high-phase-angle observations of Uranus were taken by the cameras and the IRIS instrument on the Voyager 1 and 2 spacecraft. 
The observations were taken at phase angles between 26\degr{} to 107\degr{}, but the data were limited in wavelength range and taken during Uranus' southern hemisphere summer solstice. 
The New Horizons data comparatively provide new and complementary data with a different set of wavelengths, from a better-calibrated instrument, and during a different season (approaching northern hemisphere winter solstice). 
These data, in combination with previous Voyager observations (e.g., \citealt{wenkert2023PDS}) and low-phase angle observations from Hubble and the ground could be used to assess any temporal variation in Uranus' atmosphere and more critically enable a better constraint of Uranus' albedo and energy budget in advance of a Uranus Orbiter and Probe mission \citep{NAP27209}.

While a number of understudied satellites and small bodies may have appeared to transit Uranus during the course of these observations, their brightnesses, phase functions, and orientations are uncertain. However, we can estimate the effect they might have had on the measured flux. 
For example, assuming Titania was perfectly dark (i.e., provided no additional flux) it would cause a flux decrement of approximately 0.1\% and would be below the precision of our data. Additionally, we do not predict any stars to have fallen in the background of our aperture.

Future space telescopes -- including the Nancy Grace Roman Space Telescope (Roman) with its Coronagraph Instrument \citep{mennesson2020paving} as well as the Habitable Worlds Observatory \citep{Astro2020} -- will enable direct imaging of planets in reflected light and characterization of their atmospheres.
It is critical to understand the time variability of planets to fully understand the data collected by these future observatories. 
While observations of extrasolar worlds provide only a snapshot in time, these objects are dynamic and variable over the course of their rotation and orbital periods. 
Additionally, we will not be able to image them at or near full phase due to instrumental constraints. Thus, we must use our Solar System neighbors like Uranus to provide baseline measurements for comparison to future observations.

\section{Conclusion} \label{sec:conclusion}

We have presented here a snapshot of Uranus in time as viewed by New Horizons at a phase angle of 43.9\degr{} in September 2023, 52.4\degr{} in September 2019, and 44.0\degr{} in June 2010. We have analyzed eight scans across four color filters spanning a range of wavelengths from 400 to 910\,nm. With just one rotation of Uranus from 2023 we are unable to robustly detect the presence of any features, although support imagery from both HST and amateurs suggest that few features were present at the time of the New Horizons observations. Any variation in reflectance of the New Horizons data is less than the expected error bars of the measurements.

The case of Uranus poses an interesting problem for exoplanet scientists since little will be known about an exoplanet from a single direct-imaging observation. 
Missions like the Roman Coronagraph Instrument and the Habitable Worlds Observatory will image planets in reflected light for the first time at a variety of geometries, planet-star separations, and phase angles. 
Ice giant-sized exoplanets have been found to be among the most abundant in the galaxy \citep{fulton2018AJ....156..264F}, which means it is critical that we leverage Solar System observations at the full range of phase angles to inform the atmospheric models that will be used to interpret the spectra of directly imaged exoplanets. 

\section*{acknowledgements}

This work was supported by NASA's New Horizons KEM2 Extended Mission.
This work used data from the NASA/ESA Hubble Space Telescope, and A.A.S. was supported by grants from the Space Telescope Science Institute, which is operated by the Association of Universities for Research in Astronomy, Inc., under NASA contract NAS 5-26555.  These observations are associated with program GO17294 and we thank the Space Telescope Science Institute staff for enabling early execution of the OPAL program to match the New Horizons data collection. This research has made use of the SVO Filter Profile Service ``Carlos Rodrigo", funded by MCIN/AEI/10.13039/501100011033/ through grant PID2020-112949GB-I00 \citep{2012ivoa.rept.1015R, 2020sea..confE.182R}.

We are grateful to everyone who obtained and submitted images for our public campaign. All images included the observation details. Some submissions also provided additional information, including observer names and contact information. The following observer names were received: Anthony Wesley (Rubyvale, Queensland, Australia), Blake Estes, Christian Sasse, Andy Casely (Blue Mountains), Grant Sorensen, Con Kolivas (Melbourne, Victoria, Australia), Damien, Moises Portillo Leon, Rizzi, J. Sussenbach (Houten, Utrecht, Netherlands), Raimondo Sedrani (Pordenone, Italy), Mike Wolle (Knittelfeld, Austria), Serge Dramonis (Greece), Gregory T. Shanos (Sarasota, FL, USA), Antanio, Luigi Morrone (Agerola, Italy), Frank J. Melillo (Hotsville, NY, USA), Antonio Cidadao (Carcavelos, Portugal).

\facilities{New Horizons (MVIC), HST (WFC3)}

\software{Astropy \citep{2013A&A...558A..33A,2018AJ....156..123A, 2022ApJ...935..167A}, 
        Matplotlib \citep{Hunter:2007}, Photutils \citep{bradley_2023}, NumPy \citep{harris2020array}, pandas \citep{reback2020pandas, mckinney-proc-scipy-2010}, uncertainties \citep{lebigot2016uncertainties}
        }
        
\bibliography{references}{}

\begin{thebibliography}{}
\expandafter\ifx\csname natexlab\endcsname\relax\def\natexlab#1{#1}\fi
\providecommand{\url}[1]{\href{#1}{#1}}
\providecommand{\dodoi}[1]{doi:~\href{http://doi.org/#1}{\nolinkurl{#1}}}
\providecommand{\doeprint}[1]{\href{http://ascl.net/#1}{\nolinkurl{http://ascl.net/#1}}}
\providecommand{\doarXiv}[1]{\href{https://arxiv.org/abs/#1}{\nolinkurl{https://arxiv.org/abs/#1}}}

\bibitem[{{Archinal} {et~al.}(2018){Archinal}, {Acton}, {A'Hearn}, {Conrad}, {Consolmagno}, {Duxbury}, {Hestroffer}, {Hilton}, {Kirk}, {Klioner}, {McCarthy}, {Meech}, {Oberst}, {Ping}, {Seidelmann}, {Tholen}, {Thomas}, \& {Williams}}]{archinal2018CeMDA.130...22A}
{Archinal}, B.~A., {Acton}, C.~H., {A'Hearn}, M.~F., {et~al.} 2018, Celestial Mechanics and Dynamical Astronomy, 130, 22, \dodoi{10.1007/s10569-017-9805-5}

\bibitem[{{Astropy Collaboration} {et~al.}(2013){Astropy Collaboration}, {Robitaille}, {Tollerud}, {Greenfield}, {Droettboom}, {Bray}, {Aldcroft}, {Davis}, {Ginsburg}, {Price-Whelan}, {Kerzendorf}, {Conley}, {Crighton}, {Barbary}, {Muna}, {Ferguson}, {Grollier}, {Parikh}, {Nair}, {Unther}, {Deil}, {Woillez}, {Conseil}, {Kramer}, {Turner}, {Singer}, {Fox}, {Weaver}, {Zabalza}, {Edwards}, {Azalee Bostroem}, {Burke}, {Casey}, {Crawford}, {Dencheva}, {Ely}, {Jenness}, {Labrie}, {Lim}, {Pierfederici}, {Pontzen}, {Ptak}, {Refsdal}, {Servillat}, \& {Streicher}}]{2013A&A...558A..33A}
{Astropy Collaboration}, {Robitaille}, T.~P., {Tollerud}, E.~J., {et~al.} 2013, \aap, 558, A33, \dodoi{10.1051/0004-6361/201322068}

\bibitem[{{Astropy Collaboration} {et~al.}(2018){Astropy Collaboration}, {Price-Whelan}, {Sip{\H{o}}cz}, {G{\"u}nther}, {Lim}, {Crawford}, {Conseil}, {Shupe}, {Craig}, {Dencheva}, {Ginsburg}, {VanderPlas}, {Bradley}, {P{\'e}rez-Su{\'a}rez}, {de Val-Borro}, {Aldcroft}, {Cruz}, {Robitaille}, {Tollerud}, {Ardelean}, {Babej}, {Bach}, {Bachetti}, {Bakanov}, {Bamford}, {Barentsen}, {Barmby}, {Baumbach}, {Berry}, {Biscani}, {Boquien}, {Bostroem}, {Bouma}, {Brammer}, {Bray}, {Breytenbach}, {Buddelmeijer}, {Burke}, {Calderone}, {Cano Rodr{\'\i}guez}, {Cara}, {Cardoso}, {Cheedella}, {Copin}, {Corrales}, {Crichton}, {D'Avella}, {Deil}, {Depagne}, {Dietrich}, {Donath}, {Droettboom}, {Earl}, {Erben}, {Fabbro}, {Ferreira}, {Finethy}, {Fox}, {Garrison}, {Gibbons}, {Goldstein}, {Gommers}, {Greco}, {Greenfield}, {Groener}, {Grollier}, {Hagen}, {Hirst}, {Homeier}, {Horton}, {Hosseinzadeh}, {Hu}, {Hunkeler}, {Ivezi{\'c}}, {Jain}, {Jenness}, {Kanarek}, {Kendrew}, {Kern}, {Kerzendorf}, {Khvalko}, {King}, {Kirkby}, {Kulkarni},
  {Kumar}, {Lee}, {Lenz}, {Littlefair}, {Ma}, {Macleod}, {Mastropietro}, {McCully}, {Montagnac}, {Morris}, {Mueller}, {Mumford}, {Muna}, {Murphy}, {Nelson}, {Nguyen}, {Ninan}, {N{\"o}the}, {Ogaz}, {Oh}, {Parejko}, {Parley}, {Pascual}, {Patil}, {Patil}, {Plunkett}, {Prochaska}, {Rastogi}, {Reddy Janga}, {Sabater}, {Sakurikar}, {Seifert}, {Sherbert}, {Sherwood-Taylor}, {Shih}, {Sick}, {Silbiger}, {Singanamalla}, {Singer}, {Sladen}, {Sooley}, {Sornarajah}, {Streicher}, {Teuben}, {Thomas}, {Tremblay}, {Turner}, {Terr{\'o}n}, {van Kerkwijk}, {de la Vega}, {Watkins}, {Weaver}, {Whitmore}, {Woillez}, {Zabalza}, \& {Astropy Contributors}}]{2018AJ....156..123A}
{Astropy Collaboration}, {Price-Whelan}, A.~M., {Sip{\H{o}}cz}, B.~M., {et~al.} 2018, \aj, 156, 123, \dodoi{10.3847/1538-3881/aabc4f}

\bibitem[{{Astropy Collaboration} {et~al.}(2022){Astropy Collaboration}, {Price-Whelan}, {Lim}, {Earl}, {Starkman}, {Bradley}, {Shupe}, {Patil}, {Corrales}, {Brasseur}, {N{\"o}the}, {Donath}, {Tollerud}, {Morris}, {Ginsburg}, {Vaher}, {Weaver}, {Tocknell}, {Jamieson}, {van Kerkwijk}, {Robitaille}, {Merry}, {Bachetti}, {G{\"u}nther}, {Aldcroft}, {Alvarado-Montes}, {Archibald}, {B{\'o}di}, {Bapat}, {Barentsen}, {Baz{\'a}n}, {Biswas}, {Boquien}, {Burke}, {Cara}, {Cara}, {Conroy}, {Conseil}, {Craig}, {Cross}, {Cruz}, {D'Eugenio}, {Dencheva}, {Devillepoix}, {Dietrich}, {Eigenbrot}, {Erben}, {Ferreira}, {Foreman-Mackey}, {Fox}, {Freij}, {Garg}, {Geda}, {Glattly}, {Gondhalekar}, {Gordon}, {Grant}, {Greenfield}, {Groener}, {Guest}, {Gurovich}, {Handberg}, {Hart}, {Hatfield-Dodds}, {Homeier}, {Hosseinzadeh}, {Jenness}, {Jones}, {Joseph}, {Kalmbach}, {Karamehmetoglu}, {Ka{\l}uszy{\'n}ski}, {Kelley}, {Kern}, {Kerzendorf}, {Koch}, {Kulumani}, {Lee}, {Ly}, {Ma}, {MacBride}, {Maljaars}, {Muna}, {Murphy}, {Norman},
  {O'Steen}, {Oman}, {Pacifici}, {Pascual}, {Pascual-Granado}, {Patil}, {Perren}, {Pickering}, {Rastogi}, {Roulston}, {Ryan}, {Rykoff}, {Sabater}, {Sakurikar}, {Salgado}, {Sanghi}, {Saunders}, {Savchenko}, {Schwardt}, {Seifert-Eckert}, {Shih}, {Jain}, {Shukla}, {Sick}, {Simpson}, {Singanamalla}, {Singer}, {Singhal}, {Sinha}, {Sip{\H{o}}cz}, {Spitler}, {Stansby}, {Streicher}, {{\v{S}}umak}, {Swinbank}, {Taranu}, {Tewary}, {Tremblay}, {de Val-Borro}, {Van Kooten}, {Vasovi{\'c}}, {Verma}, {de Miranda Cardoso}, {Williams}, {Wilson}, {Winkel}, {Wood-Vasey}, {Xue}, {Yoachim}, {Zhang}, {Zonca}, \& {Astropy Project Contributors}}]{2022ApJ...935..167A}
{Astropy Collaboration}, {Price-Whelan}, A.~M., {Lim}, P.~L., {et~al.} 2022, \apj, 935, 167, \dodoi{10.3847/1538-4357/ac7c74}

\bibitem[{Bradley {et~al.}(2023)Bradley, Sip{\H o}cz, Robitaille, Tollerud, Vin{\'{\i}}cius, Deil, Barbary, Wilson, Busko, Donath, G{\"u}nther, Cara, Lim, Me{\ss}linger, Conseil, Burnett, Bostroem, Droettboom, Bray, Bratholm, Jamieson, Ginsburg, Barentsen, Craig, Morris, Perrin, Rathi, Pascual, Perren, \& Georgiev}]{bradley_2023}
Bradley, L., Sip{\H o}cz, B., Robitaille, T., {et~al.} 2023, astropy/photutils: 1.10.0, 1.10.0,  Zenodo, \dodoi{10.5281/zenodo.1035865}

\bibitem[{{Calamida} {et~al.}(2022){Calamida}, {Bajaj}, {Mack}, {Marinelli}, {Medina}, {Pidgeon}, {Kozhurina-Platais}, {Shanahan}, \& {Som}}]{Calamida2022AJ....164...32C}
{Calamida}, A., {Bajaj}, V., {Mack}, J., {et~al.} 2022, \aj, 164, 32, \dodoi{10.3847/1538-3881/ac73f0}

\bibitem[{{Colina} {et~al.}(1996){Colina}, {Bohlin}, \& {Castelli}}]{Colina1996AJ....112..307C}
{Colina}, L., {Bohlin}, R.~C., \& {Castelli}, F. 1996, \aj, 112, 307, \dodoi{10.1086/118016}

\bibitem[{{Fulton} \& {Petigura}(2018)}]{fulton2018AJ....156..264F}
{Fulton}, B.~J., \& {Petigura}, E.~A. 2018, \aj, 156, 264, \dodoi{10.3847/1538-3881/aae828}

\bibitem[{{Hanel} {et~al.}(1981){Hanel}, {Conrath}, {Herath}, {Kunde}, \& {Pirraglia}}]{Hanel1981JGR....86.8705H}
{Hanel}, R., {Conrath}, B., {Herath}, L., {Kunde}, V., \& {Pirraglia}, J. 1981, \jgr, 86, 8705, \dodoi{10.1029/JA086iA10p08705}

\bibitem[{{Hanel} {et~al.}(1980){Hanel}, {Crosby}, {Herath}, {Vanous}, {Collins}, {Creswick}, {Harris}, \& {Rhodes}}]{Hanel1980ApOpt..19.1391H}
{Hanel}, R., {Crosby}, D., {Herath}, L., {et~al.} 1980, \ao, 19, 1391, \dodoi{10.1364/AO.19.001391}

\bibitem[{Harris {et~al.}(2020)Harris, Millman, van~der Walt, Gommers, Virtanen, Cournapeau, Wieser, Taylor, Berg, Smith, Kern, Picus, Hoyer, van Kerkwijk, Brett, Haldane, del R{\'{i}}o, Wiebe, Peterson, G{\'{e}}rard-Marchant, Sheppard, Reddy, Weckesser, Abbasi, Gohlke, \& Oliphant}]{harris2020array}
Harris, C.~R., Millman, K.~J., van~der Walt, S.~J., {et~al.} 2020, Nature, 585, 357, \dodoi{10.1038/s41586-020-2649-2}

\bibitem[{{Howett} {et~al.}(2017){Howett}, {Parker}, {Olkin}, {Reuter}, {Ennico}, {Grundy}, {Graps}, {Harrison}, {Throop}, {Buie}, {Lovering}, {Porter}, {Weaver}, {Young}, {Stern}, {Beyer}, {Binzel}, {Buratti}, {Cheng}, {Cook}, {Cruikshank}, {Dalle Ore}, {Earle}, {Jennings}, {Linscott}, {Lunsford}, {Parker}, {Phillippe}, {Protopapa}, {Quirico}, {Schenk}, {Schmitt}, {Singer}, {Spencer}, {Stansberry}, {Tsang}, {Weigle}, \& {Verbiscer}}]{howett2017Icar..287..140H}
{Howett}, C.~J.~A., {Parker}, A.~H., {Olkin}, C.~B., {et~al.} 2017, \icarus, 287, 140, \dodoi{10.1016/j.icarus.2016.12.007}

\bibitem[{Hunter(2007)}]{Hunter:2007}
Hunter, J.~D. 2007, Computing in Science \& Engineering, 9, 90, \dodoi{10.1109/MCSE.2007.55}

\bibitem[{{Karkoschka}(1998)}]{karkoschka1998Icar..133..134K}
{Karkoschka}, E. 1998, \icarus, 133, 134, \dodoi{10.1006/icar.1998.5913}

\bibitem[{Lebigot(2016)}]{lebigot2016uncertainties}
Lebigot, E.~O. 2016, Uncertainties: a Python package for calculations with uncertainties, 3.2.1.
\newblock \url{https://pythonhosted.org/uncertainties/}

\bibitem[{Mc{K}inney(2010)}]{mckinney-proc-scipy-2010}
Mc{K}inney, W. 2010, in {Proceedings of the 9th Python in Science Conference}, ed. {St\'efan van der Walt and Jarrod Millman }, 56 -- 61, \dodoi{10.25080/Majora-92bf1922-00a}

\bibitem[{{Meeus}(1997)}]{Meeus1997JBAA..107..332M}
{Meeus}, J. 1997, Journal of the British Astronomical Association, 107, 332

\bibitem[{{Mennesson} {et~al.}(2020){Mennesson}, {Juanola-Parramon}, {Nemati}, {Ruane}, {Bailey}, {Bolcar}, {Martin}, {Zimmerman}, {Stark}, {Pueyo}, {Benford}, {Cady}, {Crill}, {Douglas}, {Gaudi}, {Kasdin}, {Kern}, {Krist}, {Kruk}, {Luchik}, {Macintosh}, {Mandell}, {Mawet}, {McEnery}, {Meshkat}, {Poberezhskiy}, {Rhodes}, {Riggs}, {Turnbull}, {Roberge}, {Shi}, {Siegler}, {Stapelfeldt}, {Ygouf}, {Zellem}, \& {Zhao}}]{mennesson2020paving}
{Mennesson}, B., {Juanola-Parramon}, R., {Nemati}, B., {et~al.} 2020, arXiv e-prints, arXiv:2008.05624, \dodoi{10.48550/arXiv.2008.05624}

\bibitem[{{National Academies of Sciences Engineering and Medicine}(2021)}]{Astro2020}
{National Academies of Sciences Engineering and Medicine}. 2021, {Pathways to Discovery in Astronomy and Astrophysics for the 2020s} (Washington, D.C.: National Academies Press), 616, \dodoi{10.17226/26141}

\bibitem[{{National Academies of Sciences, Engineering, and Medicine}(2023)}]{NAP27209}
{National Academies of Sciences, Engineering, and Medicine}. 2023, {Origins, Worlds, and Life: Planetary Science and Astrobiology in the Next Decade} (Washington, DC: The National Academies Press), \dodoi{10.17226/27209}

\bibitem[{pandas~development team(2020)}]{reback2020pandas}
pandas~development team, T. 2020, pandas-dev/pandas: Pandas, 2.2.2,  Zenodo, \dodoi{10.5281/zenodo.3509134}

\bibitem[{Pearl \& Conrath(1991)}]{Pearl1991-xv}
Pearl, J.~C., \& Conrath, B.~J. 1991, J. Geophys. Res., 96, 18921, \dodoi{10.1029/91ja01087}

\bibitem[{Pearl {et~al.}(1990)Pearl, Conrath, Hanel, Pirraglia, \& Coustenis}]{Pearl1990-od}
Pearl, J.~C., Conrath, B.~J., Hanel, R.~A., Pirraglia, J.~A., \& Coustenis, A. 1990, Icarus, 84, 12, \dodoi{10.1016/0019-1035(90)90155-3}

\bibitem[{Peterson {et~al.}(2008)Peterson, Versteeg, Parker, Steffl, McCabe, Lunsford, Weaver, Taylor, Olkin, Livi, Carcich, James, \& Elliott}]{Peterson-2008-NHSOC}
Peterson, J., Versteeg, M., Parker, J., {et~al.} 2008, {New Horizons SOC to Instrument Pipeline ICD, Document 05310-SOCINST-01}, 1.0,  Southwest Research Institute

\bibitem[{{Reuter} {et~al.}(2008){Reuter}, {Stern}, {Scherrer}, {Jennings}, {Baer}, {Hanley}, {Hardaway}, {Lunsford}, {McMuldroch}, {Moore}, {Olkin}, {Parizek}, {Reitsma}, {Sabatke}, {Spencer}, {Stone}, {Throop}, {van Cleve}, {Weigle}, \& {Young}}]{Reuter2008SSRv..140..129R}
{Reuter}, D.~C., {Stern}, S.~A., {Scherrer}, J., {et~al.} 2008, \ssr, 140, 129, \dodoi{10.1007/s11214-008-9375-7}

\bibitem[{{Rodrigo} \& {Solano}(2020)}]{2020sea..confE.182R}
{Rodrigo}, C., \& {Solano}, E. 2020, in XIV.0 Scientific Meeting (virtual) of the Spanish Astronomical Society, 182

\bibitem[{{Rodrigo} {et~al.}(2012){Rodrigo}, {Solano}, \& {Bayo}}]{2012ivoa.rept.1015R}
{Rodrigo}, C., {Solano}, E., \& {Bayo}, A. 2012, {SVO Filter Profile Service Version 1.0}, IVOA Working Draft 15 October 2012, \dodoi{10.5479/ADS/bib/2012ivoa.rept.1015R}

\bibitem[{Simon(2015)}]{HST_OPAL_DOI}
Simon, A. 2015, Outer Planet Atmospheres Legacy (``OPAL"),  STScI/MAST, \dodoi{10.17909/T9G593}

\bibitem[{{Simon} {et~al.}(2015){Simon}, {Wong}, \& {Orton}}]{Simon2015ApJ...812...55S}
{Simon}, A.~A., {Wong}, M.~H., \& {Orton}, G.~S. 2015, \apj, 812, 55, \dodoi{10.1088/0004-637X/812/1/55}

\bibitem[{{Smith} {et~al.}(1986){Smith}, {Soderblom}, {Beebe}, {Bliss}, {Boyce}, {Brahic}, {Briggs}, {Brown}, {Collins}, {Cook}, {Croft}, {Cuzzi}, {Danielson}, {Davies}, {Dowling}, {Godfrey}, {Hansen}, {Harris}, {Hunt}, {Ingersoll}, {Johnson}, {Krauss}, {Masursky}, {Morrison}, {Owen}, {Plescia}, {Pollack}, {Porco}, {Rages}, {Sagan}, {Shoemaker}, {Sromovsky}, {Stoker}, {Strom}, {Suomi}, {Synnott}, {Terrile}, {Thomas}, {Thompson}, \& {Veverka}}]{smith1986Sci...233...43S}
{Smith}, B.~A., {Soderblom}, L.~A., {Beebe}, R., {et~al.} 1986, Science, 233, 43, \dodoi{10.1126/science.233.4759.43}

\bibitem[{{Wenkert} {et~al.}(2022){Wenkert}, {Deen}, \& {Bunch}}]{wenkert2022AGUFM.P32E1865W}
{Wenkert}, D., {Deen}, R.~G., \& {Bunch}, W.~L. 2022, in AGU Fall Meeting Abstracts, Vol. 2022, P32E--1865

\bibitem[{{Wenkert}(2023)}]{wenkert2023PDS}
{Wenkert}, D.~D. 2023, {Restoring and Archiving Voyager 1 Cruise Images of Uranus and Neptune (RAV1CIUN) PDART Bundle},  Planetary Data System, \dodoi{10.17189/t2r8-rk88}

\end{thebibliography}
\bibliographystyle{aasjournal}

\end{document}